\documentclass[twocolumn,showkeys,aps,prd,showpacs]{revtex4-1}
\UseRawInputEncoding
\usepackage{graphicx}
\usepackage[CJKbookmarks,dvipdfm,colorlinks,linkcolor=blue,citecolor=blue]{hyperref}
\usepackage{amsmath, amssymb}
\usepackage{makecell}
\usepackage{multirow}
\usepackage{CJK}
\usepackage{mathrsfs}
\usepackage{bm}
\usepackage{amsmath}
\usepackage{dcolumn}
\usepackage{epstopdf}
\usepackage{dsfont}
\usepackage{amssymb}
\usepackage{tabularx}
\usepackage{array}
\usepackage{float}
\usepackage{color}
\usepackage{epstopdf}
\usepackage{mathrsfs}
\usepackage{extarrows}

\begin{document}

\title{Anomalous valley Hall effect in electric-potential-difference antiferromagnetic $\mathrm{Cr_2CHCl}$ monolayer}
\author{Dun-Cheng Liang and San-Dong Guo}
\email{sandongyuwang@163.com}
\affiliation{School of Electronic Engineering, Xi'an University of Posts and Telecommunications, Xi'an 710121, China}
\author{Shaobo Chen}
\affiliation{College of Electronic and Information Engineering, Anshun University, Anshun 561000, People¡¯s Republic of China}

\begin{abstract}
The antiferromagnetic (AFM) valleytronics  can be intrinsically more energy-saving and fast-operating in
device applications.  In general, the lacking spontaneous spin-splitting hinders the implementation and detection of anomalous valley Hall effect (AVHE). Here, we propose  to implement AVHE in electric-potential-difference antiferromagnetic $\mathrm{Cr_2CHCl}$ monolayer with excellent stability, where the spontaneous spin-splitting can be induced due to   layer-dependent electrostatic potential
caused by out-of-plane built-in electric field.  From a symmetry perspective,  the introduction of Janus structure  breaks the combined symmetry ($PT$ symmetry) of spatial
inversion ($P$) and time reversal ($T$), which gives rise to spin-splitting. Both unstarined and strained monolayer $\mathrm{Cr_2CHCl}$ possess valley splitting of larger than  51 meV, which is higher than
the thermal energy of  room temperature (25 meV).
The layer-locked  Berry curvature  gives rise to  layer-locked AVHE.
Our work reveals a route to achieve
AVHE in AFM monolayer with spontaneous spin-splitting.

\end{abstract}

\maketitle

\section{Introduction}
 The discovery and successful preparation of rich two-dimensional (2D)
materials lays the foundation for  valleytronics, which can process information and perform  logic operations  with low power consumption and high speed\cite{q1,q2,q3,q4,ref1,ref2,ref3,ref4}. Recently, the ferrovalley semiconductor (FVS)  has been proposed to  realize intrinsic valley polarization\cite{q10}, which appears in out-of-plane hexagonal ferromagnetic (FM)
materials with broken spatial inversion symmetry\cite{q10,q11,q12,q13,q13-1,q14,q15,q16,q17,q18}. These  offer interesting platforms to study valley-contrasting transport and Berry physics. In the FVS, valley-dependent Berry curvature can produce anomalous valley Hall effect (AVHE), where the charge Hall current originates from the spontaneous valley polarization.
The antiferromagnetic
(AFM) materials possess  the high storage density, robustness against external magnetic field, as well as the ultrafast writing speed\cite{v12}, so realizing  valley polarization and AVHE in AFM materials is more meaningful for valleytronic application.

For 2D AFM materials with $PT$ symmetry (a combined symmetry  of spatial
inversion ($P$) and time reversal ($T$)), there is zero berry curvature ($\Omega(k)$)  and no spin-splitting everywhere in the momentum space,  which hinders the realization of AVHE.  Recently,  an intuitive way is proposed  to produce spin-splitting in AFM materials by making the magnetic atoms with opposite spin polarization locating in the different environment (surrounding atomic arrangement)\cite{gsd}.
The altermagnetism\cite{k6} and electric-potential-difference antiferromagnetism (EPD-AFM)\cite{k7} are the representative examples, and they possess intrinsic  spin-splitting. For  altermagnetism, the two different environments (The surrounding atoms are in the same arrangement, yet not in the same orientation.) can be connected by special symmetry operation\cite{k6}. For EPD-AFM, the different  environments occupied by two (spin-up and spin-down) magnetic atoms are due to an electric-potential-difference caused by   an out-of-plane built-in electric field, and the magnetic atoms have opposite layer spin polarization (A-type AFM ordering). If EPD-AFM has hexagonal symmetry, the layer-locked Berry curvature can appear\cite{gsd1}. If hexagonal symmetry couples  with the out-of-plane magnetization for EPD-AFM, spontaneous valley polarization will exist\cite{gsd1}. Therefore, an out-of-plane hexagonal EPD-AFM with  energy extrema of conduction or
valance bands located at high symmetry -K and K points can achieve AVHE.

The experimentally synthesized $\mathrm{Cr_2C}$  is a half-metallic
ferromagnet\cite{v16}.
 By surface functionalization  with
H or Cl in $\mathrm{Cr_2C}$,   both  ferromagnetic-antiferromagnetic transition and
metal-insulator transition  can be induced simultaneously\cite{v15}. The functionalized $\mathrm{Cr_2CH_2}$ and $\mathrm{Cr_2CCl_2}$ possess A-type AFM ordering and energy extrema at  -K and K high symmetry points. However, for $\mathrm{Cr_2CH_2}$ and $\mathrm{Cr_2CCl_2}$,   the spin degeneracy of -K and K valleys is maintained  due to $PT$ symmetry, which prohibits the AVHE. Based on $\mathrm{Cr_2CH_2}$ and $\mathrm{Cr_2CCl_2}$, a Janus $\mathrm{Cr_2CHCl}$ can be constructed to achieve hexagonal EPD-AFM, which is a possibe candidate material to achieve AVHE.

\begin{figure}
  \includegraphics[width=8cm]{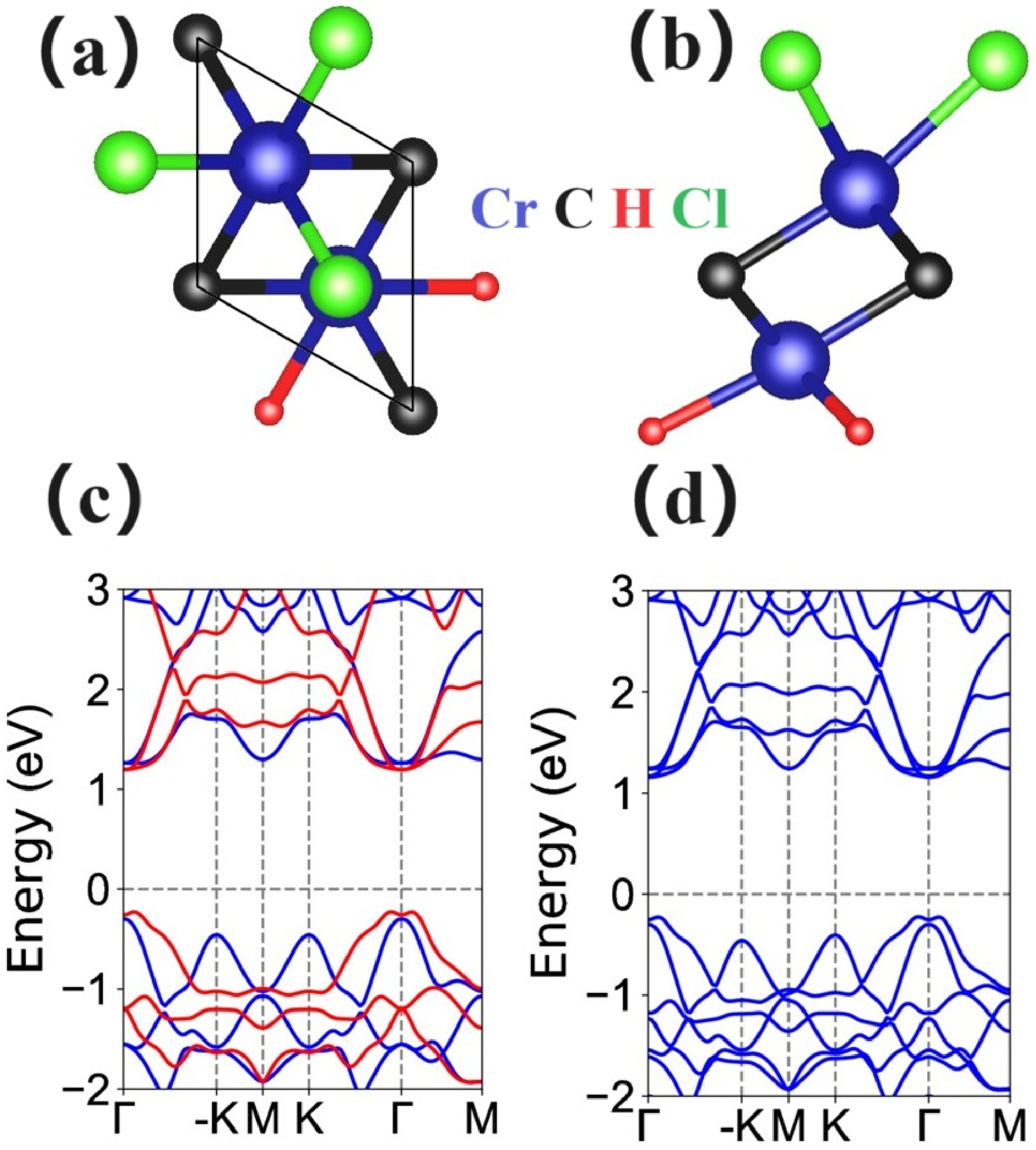}
  \caption{(Color online) For  monolayer  $\mathrm{Cr_2CHCl}$,  the top (a) and side (b) views of crystal structures;  the energy band structures without (c) and  with (d) SOC. In (c), the spin-up
and spin-down channels are depicted in blue and red.}\label{st}
\end{figure}

\section{Computational detail}
Within density functional theory (DFT)\cite{1}, we perform the spin-polarized  first-principles calculations  within the projector augmented-wave (PAW) method,  as implemented in Vienna ab initio Simulation Package (VASP)\cite{pv1,pv2,pv3}.  We use the generalized gradient
approximation  of Perdew-Burke-Ernzerhof (PBE-GGA)\cite{pbe}as the exchange-correlation functional.
The kinetic energy cutoff  of 500 eV,  total energy  convergence criterion of  $10^{-8}$ eV, and  force convergence criterion of 0.0001 $\mathrm{eV.{\AA}^{-1}}$  are adopted. To account for the localized nature of Cr-3$d$ orbitals, a Hubbard correction $U_{eff}$=3.0 eV\cite{v13,v13-1} is used by  the
rotationally invariant approach proposed by Dudarev et al\cite{u}. The spin-orbital coupling (SOC) is included to investigate valley splitting and magnetic anisotropy energy (MAE).
To avoid interactions between neighboring slabs, the vacuum space of more than 20 $\mathrm{{\AA}}$ along $z$ direction is added.
We use a 21$\times$21$\times$1 Monkhorst-Pack k-point meshes to sample the Brillouin zone (BZ) for calculating electronic structures.
Based on  finite displacement method with 5$\times$5$\times$1 supercell, the interatomic force constants (IFCs)  are calculated  with AFM ordering, and the phonon dispersion is constructed by the  Phonopy code\cite{pv5}.
The ab initio
molecular dynamics (AIMD) simulations  using NVT ensemble are performed with a 4$\times$4$\times$1 supercell  for more than
8000 fs with a time step of 1 fs.
The Berry curvatures
are calculated directly from the calculated
wave functions  based on Fukui's
method\cite{bm},  as implemented in  the VASPBERRY code\cite{bm1,bm2,bm3}.

\begin{figure}
  \includegraphics[width=8cm]{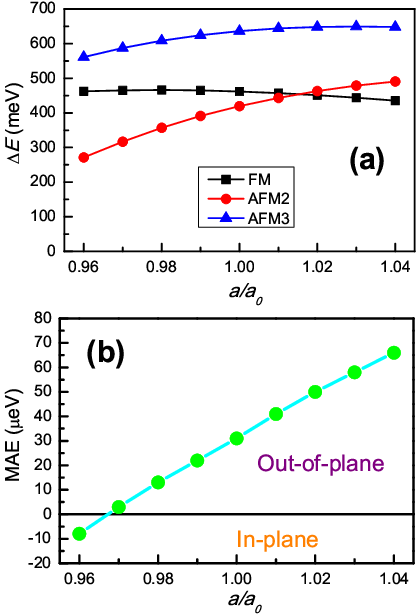}
\caption{(Color online)For $\mathrm{Cr_2CHCl}$, the energy differences per unit cell between FM/AFM2/AFM3 and AFM1 (a) and MAE (b)  as a function of $a/a_0$. }\label{emae}
\end{figure}
\begin{figure*}
   \includegraphics[width=16cm]{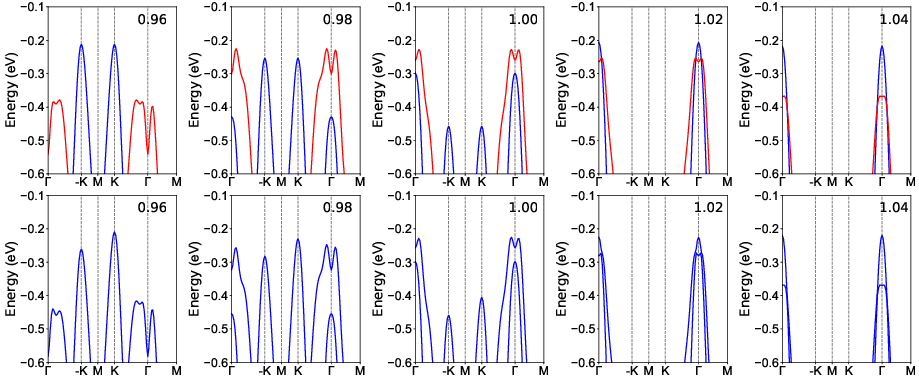}
\caption{(Color online)For  $\mathrm{Cr_2CHCl}$, the energy band structures of valence bands near the Fermi level without (top plane) and  with (bottom plane) SOC at representative $a/a_0$ (0.96, 0.98, 1.00, 1.02 and 1.04). For top plane, the blue (red) lines represent the band structure in the spin-up (spin-down) direction.}\label{band-1}
\end{figure*}

\begin{figure}
  \includegraphics[width=8cm]{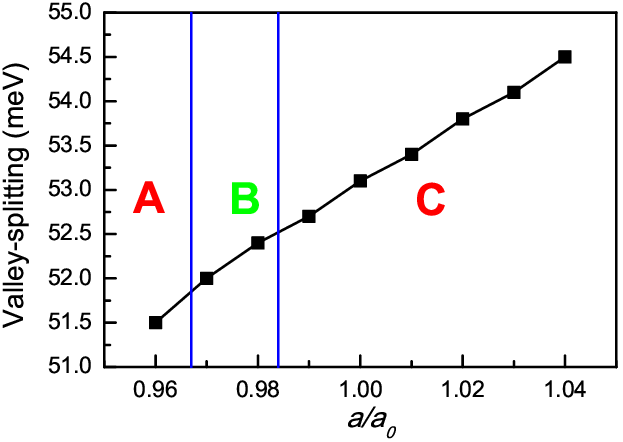}
  \caption{(Color online)For  $\mathrm{Cr_2CHCl}$ with assumed out-of-plane  magnetization direction , the valley splitting as a function of $a/a_0$.  For region A,  no spontaneous valley polarization appears due to the in-plane magnetization direction. For region C, the VBM is not at K/-K point. For region B, it is suitable to produce AVHE. }\label{vv}
\end{figure}

\begin{figure*}
  \includegraphics[width=14cm]{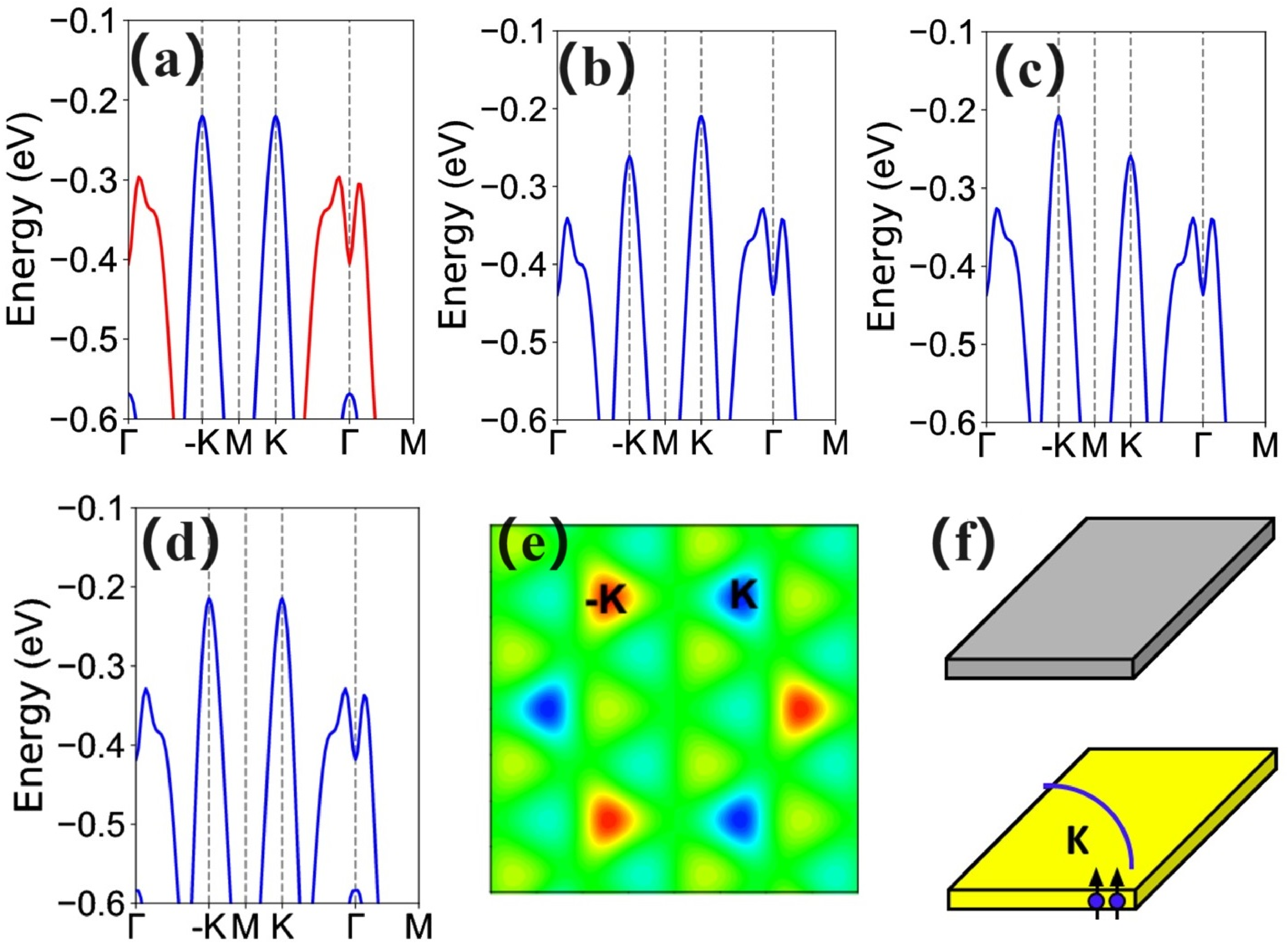}
\caption{(Color online)For  $\mathrm{Cr_2CHCl}$  with $a/a_0$=0.97, the energy band structures  without SOC (a), and  with SOC (b, c, d) for magnetization direction along the positive $z$, negative $z$, and positive $x$ direction, respectively; the distribution of Berry curvatures of    spin-up (e); in the presence of a longitudinal in-plane electric field,  an appropriate hole doping  produces  layer-locked AVHE (f), and  the upper and lower planes represent  the top and bottom Cr layers.  In (a), the blue (red) lines represent the band structure in the spin-up (spin-down) direction.}\label{band-2}
\end{figure*}

\section{Crystal and electronic structures}
 The $\mathrm{Cr_2CH_2}$ and $\mathrm{Cr_2CCl_2}$  monolayers  have been proved to  possess A-type AFM ordering with  good stabilities\cite{v15}. Monolayer $\mathrm{Cr_2CHCl}$ can be constructed by  replacing one of two  H (Cl)  layers with Cl (H)  atoms in  $\mathrm{Cr_2CCH_2}$ ($\mathrm{Cr_2CCl_2}$) monolayer.
 The crystal structures of $\mathrm{Cr_2CHCl}$ are  shown in \autoref{st} (a) and (b), which crystallizes in the  $P3m1$ (No.~156),  lacking spatial  inversion symmetry. The $\mathrm{Cr_2CHCl}$ consists of five atomic layers in the sequence of H-Cr-C-Cr-Cl, which can produce built-in electric field due to special Janus structure. In addition, the magnetic Cr atoms of $\mathrm{Cr_2CHCl}$ distribute in two layers. These provide the basic conditions for EPD-AFM.   The FM and three  AFM configurations (AFM1, AFM2 and AFM3) are constructed, as  shown in FIG.S1 of electronic supplementary information (ESI), to determine magnetic ground state of  $\mathrm{Cr_2CHCl}$.  The AFM1 ordering is called A-type AFM state, which is necessary to form EPD-AFM.
  It is found that the energy of AFM1 per unit cell is 462 meV, 419 meV and  636 meV  lower  than those of FM, AFM2 and AFM3 cases within GGA+$U$, confirming that the monolayer  $\mathrm{Cr_2CHCl}$  possesses AFM1 ground state with the optimized  equilibrium lattice constants $a$=$b$=3.09 $\mathrm{{\AA}}$. The calculated phonon spectrums  of   $\mathrm{Cr_2CHCl}$  show no obvious imaginary frequencies (see FIG.S2 of ESI),  indicating its dynamic stability. Based on FIG.S3 of ESI, the AIMD simulation shows that  the framework of $\mathrm{Cr_2CHCl}$ is well preserved with   little fluctuations of total
energy with increasing time during the
simulation period,   which confirms its thermal stability.

The  magnetization direction is  of great significance for generating spontaneous valley polarization, and only out-of-plane magnetization direction can produce spontaneous valley splitting\cite{gsd1}.   The magnetic orientation can be  determined by calculating MAE, which can be calculated  by $E_{MAE}=E^{||}_{SOC}-E^{\perp}_{SOC}$.  The $E^{||}_{SOC}$ and $E^{\perp}_{SOC}$ are the energies that spins lie in-plane and out-of-plane, respectively. The calculated MAE is 31$\mathrm{\mu eV}$/unit cell, and the positive value indicates the out-of-plane easy magnetization axis of $\mathrm{Cr_2CHCl}$, which means  spontaneous valley polarization.

 The energy band structures of $\mathrm{Cr_2CHCl}$  are plotted in \autoref{st} (c) and (d) without SOC and  with SOC for magnetization direction along the positive $z$ direction.
According to \autoref{st} (c), the obvious
spin-splitting can be observed due to the broken $PT$ symmetry, and $\mathrm{Cr_2CHCl}$ is  an indirect band
gap semiconductor. For $\mathrm{Cr_2CHCl}$, there are -K and K valleys with  energy degeneracy in the valence band,  but the valence band maximum  (VBM) is not at the -K/K high symmetry point. The spin-splitting of $\mathrm{Cr_2CHCl}$ stems from a layer-dependent electrostatic potential, which occurs across the entire momentum space, producing  the $s$-wave symmetry of spin-splitting\cite{k7}. This is  different from altermagnetism ($d$, $g$, $i$-wave symmetry of spin-splitting) with momentum dependent spin-splitting\cite{k6}.
When SOC is included,   the  spontaneous valley polarization with the valley splitting of 53 meV ($\Delta E_V=E_{K}^V-E_{-K}^V$) can be observed (\autoref{st} (d)), and the energy of K valley
is higher than one of -K valley.
For $\mathrm{Cr_2CHCl}$, the  total magnetic moment per unit cell is strictly 0.00 $\mu_B$, and  the   magnetic moment of bottom/top Cr atom  is 3.13 $\mu_B$/-3.06 $\mu_B$.  For EPD-AFM, the two types of magnetic (up and down) atoms cannot be connected by mirror or rotation symmetry, and the absolute values of their magnetic moments  are not strictly equal, which is different from  altermagnetism with strictly equal magnetic moments for up and down atoms\cite{k6}.

\begin{figure}
   \includegraphics[width=8cm]{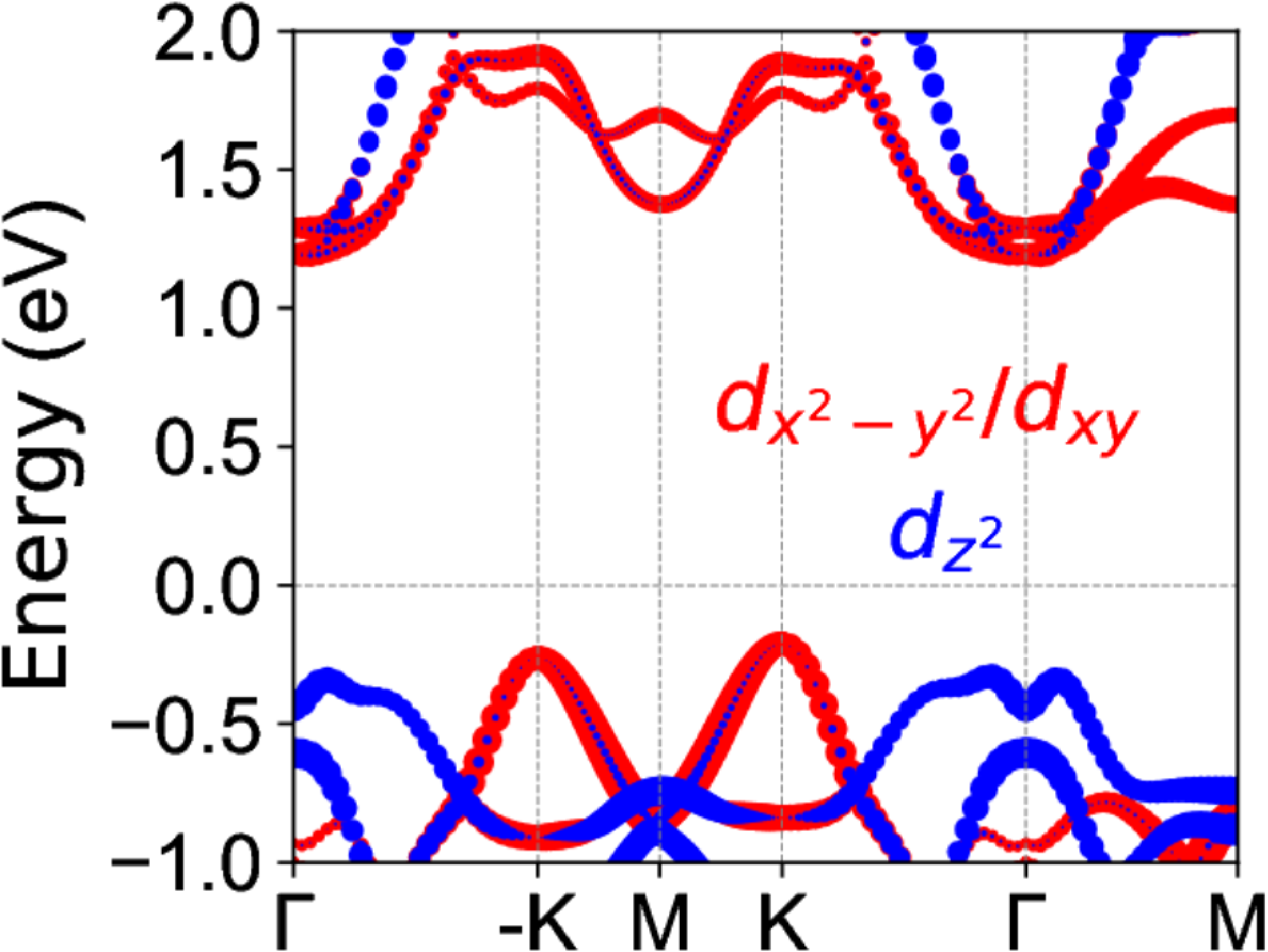}
  \caption{(Color online)For  $\mathrm{Cr_2CHCl}$  with $a/a_0$=0.97, Cr-$d$-orbital characters of energy bands within SOC.}\label{p}
\end{figure}

\section{strain effects}
The VBM of $\mathrm{Cr_2CHCl}$ is not at the -K/K high symmetry point, which is not conducive to implementation of AVHE.
Strain is a very effective way to tune the position of energy extrema for 2D materials\cite{gsd2}. Here, the biaxial strain is applied to make VBM of  $\mathrm{Cr_2CHCl}$  become -K or K point. We use $a/a_0$ (0.96 to 1.04)  to simulate the  biaxial strain, and  the $a/a_0$$<$1 ($a/a_0$$>$1) means the compressive (tensile) strain, where   $a$ and $a_0$ are the strained and  unstrained lattice constants.
Firstly,  the magnetic ground state of strained $\mathrm{Cr_2CHCl}$  is determined by calculating energy differences ($\Delta E$) between  FM/AFM2/AFM3  and  AFM1 states  vs $a/a_0$, as shown in \autoref{emae} (a).
Within considered strain range,  the  positive $\Delta E$  confirm the AFM1 ground state of strained $\mathrm{Cr_2CHCl}$.

The energy band structures of $\mathrm{Cr_2CHCl}$ at representative $a/a_0$ without SOC and with SOC (out-of-plane magnetization direction)  are shown in FIG.S4 of ESI, and the  enlarged figures of the valence band  near the Fermi level  are plotted in \autoref{band-1}.
For  valence  band,  the valley splitting   as a function of $a/a_0$ are plotted in \autoref{vv}.
It is clearly seen that the -K and K valleys always exist in the valence band, and they are from spin-up channel.  It is found that the compressive strain can make -K/K valley become VBM, and the critical point is approximately 0.984 for $a/a_0$.
The  valley splitting  of larger than 51 meV is maintained within considered $a/a_0$ range, and no valley polarization transition is produced. This is different from Janus $\mathrm{GdClF}$ monolayer\cite{gsd3}, where  the valley polarization transition can be driven  by biaxial strain.

To achieve spontaneous valley splitting, another key factor is  out-of-plane magnetization of strained $\mathrm{Cr_2CHCl}$.
 The MAE as a function of $a/a_0$ is  plotted in \autoref{emae} (b). It is found that compressive strain can induce the transition of magnetization direction, and the  magnetization of  $\mathrm{Cr_2CHCl}$ changes from out-of-plane case to in-plane case, when the $a/a_0$ is lower than 0.968.
  By considering the MAE and position of VBM together,   $\mathrm{Cr_2CHCl}$  is suitable to produce AVHE, when the $a/a_0$ is between 0.968 and 0.984.

Taking  $a/a_0$=0.97 as a example, the energy band structures of  valence bands near the Fermi energy level without SOC and with  SOC for magnetization direction along the positive $z$, negative $z$, and positive $x$ directions are shown in \autoref{band-2} (a), (b), (c) and (d).
\autoref{band-2} (a) shows obvious spin-splitting, and the VBM is at -K/K valley form the spin-up channel. However, no valley splitting can be observed. \autoref{band-2} (b) shows the  the valley splitting of 52 meV, which is higher than
the thermal energy of  room temperature (25 meV). The energy of K valley
is higher than one of -K valley. \autoref{band-2} (c)  indicates that the valley polarization transition can  be
 achieved by reversing the magnetization direction.  \autoref{band-2} (d)  shows that the in-plane magnetization direction  produces no valley polarization.

The SOC-induced valley splitting  is mainly from the intra-atomic interaction  $\hat{H}^0_{SOC}$ (the interaction
between the same spin states)\cite{q18}:
\begin{equation}\label{e1}
\hat{H}^0_{SOC}=\lambda\hat{S}_{z^`}(\hat{L}_z cos\theta+\frac{1}{2}\hat{L}_{+}e^{-i\phi}sin\theta+\frac{1}{2}\hat{L}_{-}e^{+i\phi}sin\theta)
\end{equation}
  in which $\hat{S}$, $\hat{L}$ and  $\theta$/$\phi$ mean the spin angular momentum,  orbital angular momentum  and the polar angles of spin orientation, respectively.
  For $d_{x^2-y^2}$/$d_{xy}$-dominated -K/K valley with the group symmetry of $C_{3h}$, the valley splitting $\Delta E_V=4\alpha cos\theta$\cite{q18}, where the $\alpha$ is $\lambda|\hat{S}_z|$, and the $\theta$=0/90$^{\circ}$ means out-of-plane/in-plane direction.
 For  $\mathrm{Cr_2CHCl}$  with $a/a_0$=0.97, the Cr-$d$-orbital characters of energy bands within SOC are plotted in \autoref{p}, which shows $d_{x^2-y^2}$/$d_{xy}$-orbital-dominated -K and K valleys of valence bands. When the magnetization direction  is along out-of-plane/in-plane case, the valley splitting of  $\mathrm{Cr_2CHCl}$ will be 4$\alpha$/0.

Due to A-type AFM ordering in $\mathrm{Cr_2CHCl}$,  the layer-locked Berry curvature can be observed\cite{gsd1}. Because -K and K valleys are from spin-up channel, the distribution of Berry curvatures of    spin-up is shown  in \autoref{band-2} (e).  It is clearly seen that the  Berry curvatures are opposite  for -K and K valleys. By applying a longitudinal in-plane electric field,
the Bloch carriers will acquire an anomalous transverse
velocity $v_{\bot}$$\sim$$E_{\parallel}\times\Omega(k)$\cite{v17}.
 When the Fermi level is shifted between the -K and K valleys of  the valence band, only the spin-up holes of K valley move to the bottom boundary of the sample under an in-plane electric field (\autoref{band-2} (f)), producing layer-locked AVHE.
   This accumulation of spin-polarized holes produces  a net charge/spin current, which will generate observable voltage.

\section{Conclusion}
In summary,  we present a  hexagonal AFM monolayer $\mathrm{Cr_2CHCl}$ with  spontaneous spin-splitting to realize AVHE.
 The spontaneous valley polarization can  occur in $\mathrm{Cr_2CHCl}$ with the valley splitting of larger than 51 meV due to intrinsic  out-of-plane magnetization, when the $a/a_0$ is larger than 0.968.  The introduction of
a built-in electric field  caused by Janus structure induces the  spin-splitting  in monolayer $\mathrm{Cr_2CHCl}$ due to layer-dependent electrostatic potential.  By combining with layer-locked  Berry curvature, the layer-locked AVHE can be achieved in monolayer $\mathrm{Cr_2CHCl}$ without applying out-of-plane external electric field.
Our works enrich AFM valleytronic materials, and  provide  advantageous for the development of energy-efficient and ultrafast electronic devices.

\begin{acknowledgments}
This work is supported by Natural Science Basis Research Plan in Shaanxi Province of China  (2021JM-456). We are grateful to Shanxi Supercomputing Center of China, and the calculations were performed on TianHe-2.
\end{acknowledgments}

\end{document}